\documentclass[aps,prb,twocolumn,superscriptaddress,10pt,showkeys]{revtex4-1}

\usepackage{graphicx}
\usepackage{color}
\usepackage{amssymb,amsmath}
\usepackage[hidelinks=true]{hyperref}

\newcommand{\be}{\begin{equation}}
\newcommand{\ee}{\end{equation}}
\newcommand{\bea}{\begin{eqnarray}}
\newcommand{\eea}{\end{eqnarray}}

\newcommand{\p}{\partial}

\newcommand{\lp}{\left(}
\newcommand{\rp}{\right)}

\renewcommand{\vec}[1]{{\bf #1}}

\begin{document}

\title{
Nonlocal Response and Anamorphosis: the Case of Few-Layer Black Phosphorus
}

\author{A. Mishchenko}
\thanks{These authors contributed equally.}
\affiliation{School of Physics and Astronomy, University of Manchester, Manchester M13 9PL, UK}
\author{Y. Cao}
\thanks{These authors contributed equally.}
\affiliation{School of Physics and Astronomy, University of Manchester, Manchester M13 9PL, UK}
\affiliation{Centre for Mesoscience and Nanotechnology, University of Manchester, Manchester M13 9PL, UK}
\author{G. L. Yu}
\thanks{These authors contributed equally.}
\affiliation{School of Physics and Astronomy, University of Manchester, Manchester M13 9PL, UK}
\author{C. R. Woods}
\affiliation{School of Physics and Astronomy, University of Manchester, Manchester M13 9PL, UK}
\author{R.~V.~Gorbachev}
\affiliation{School of Physics and Astronomy, University of Manchester, Manchester M13 9PL, UK}
\affiliation{Centre for Mesoscience and Nanotechnology, University of Manchester, Manchester M13 9PL, UK}
\author{K. S. Novoselov}
\affiliation{School of Physics and Astronomy, University of Manchester, Manchester M13 9PL, UK}
\author{A. K. Geim}
\affiliation{School of Physics and Astronomy, University of Manchester, Manchester M13 9PL, UK}
\affiliation{Centre for Mesoscience and Nanotechnology, University of Manchester, Manchester M13 9PL, UK}
\author{L. S. Levitov}
\email[To whom correspondence should be addressed: ]{artem.mishchenko@gmail.com, levitov@mit.edu}
\affiliation{Department of Physics, Massachusetts Institute of Technology, Cambridge, MA 02139, USA}
%%%%%%%%%%%%%%%%%%%%%%%%%%%%%%%%%%%%%%%%%%%%%%%%%%%%%%%%%%%%%%%%%%%%%%%%%%%%%%%%%%%%%%%%%%%%%%%

\begin{abstract}
\textbf{Abstract  $|$}
Few-layer black phosphorus was recently rediscovered as a narrow-bandgap atomically thin semiconductor and has already attracted unprecedented attention due to its interesting properties. One feature of this material that sets it apart from other atomically thin crystals is its structural in-plane anisotropy which manifests in strongly anisotropic transport characteristics. However, traditional angle-resolved conductance measurements present a challenge for nanoscale systems such as black phosphorus, calling for new approaches in precision studies of transport anisotropy. Here we show that the nonlocal response, being exponentially sensitive to the anisotropy value, provides a powerful tool for determining the anisotropy. This is established by combining  measurements of  the orientation-dependent nonlocal resistance response with the analysis based on the {\it anamorphosis} relations. We demonstrate that the nonlocal response can differ by orders of magnitude for different crystallographic directions even when the anisotropy is at most order-one, allowing us to extract accurate anisotropy values.
\end{abstract}

%%%%%%%%%%%%%%%%%%%%%%%%%%%%%%%%%%%%%%%%%%%%%%%%%%%%%%%%%%%%%%%%%%%%%%%%%%%%%%%%%%%%%%%%%%%%%%%



\maketitle

Black phosphorus (BP), a layered crystal allotrope of phosphorus, is composed of corrugated sheets of phosphorene held together by weak van der Waals interactions.\cite{Brown1965} In phosphorene sheets, each phosphorus atom covalently bonds to its three nearest neighbors, forming warped hexagons in a non-planar, chair conformation. This \textit{sp}${}^3$-type bonding results in a distinctly anisotropic crystal structure: phosphorene adopts an accordion-like folded hexagonal lattice with grooves running along the $y$ axis (with zigzag termination). Such crystal structure leads to highly anisotropic energy bands:\cite{Takao1981} electrons and holes of phosphorene are nearly an order of magnitude heavier in the $y$ direction (zigzag) as compared to the $x$ direction (armchair). This results in an interesting  in-plane anisotropy in electronic, optical and acoustic properties.\cite{Ling2015} Further, electrical conductivity and carrier mobility are higher along the $x$ axis than along the $y$ axis, whereas the opposite anisotropy sign is predicted for thermal conductance, a property that can potentially lead to high-efficiency thermoelectric devices.\cite{Fei2014a} The in-plane anisotropy is also of interest for plasmonics, since the plasmon resonance frequency could be tuned with the light polarization due to the highly anisotropic nature of plasmons in BP.\cite{Low2014a} However, while being of high interest because of their promise for device research and engineering, the anisotropic properties are currently lacking reliable characterization techniques. Here we develop a new approach that is particularly well suited for nanosystems where other known techniques can perform poorly or lack precision.
\par 
Transport anisotropy was studied recently in a two-terminal configuration by angle-resolved conductance measurements.\cite{Xia2014,Liu2014} Refs.~\onlinecite{Xia2014,Liu2014} employed a circular geometry, measuring two-terminal conductance between sets of opposite contacts spaced by $30^{\rm o}$ or $45^{\rm o}$. The angular dependence for the two-terminal conductance was fitted to an expression 
$\sigma(\theta)=\sigma_{xx}\cos^2⁡\theta+\sigma_{yy}\sin^2\theta$, with $\theta$ the angle between the chosen pair of opposite contacts and the $x$ axis. From the measured dependence, both the crystallographic orientation and the values $\sigma_{xx}$, $\sigma_{yy}$ could be inferred, giving the anisotropy value 
\be\label{eq:A_defined}
A = \sigma_{xx}/\sigma_{yy}
\ee 
ranging from $1.5$ to $1.8$. Unfortunately, such direct two-terminal measurements are prone to inaccuracy due to contact resistance and variability in device geometry, and thus are expected to yield relatively large errors in the inferred anisotropy values.
\par
Here we employ a different approach. We obtain the in-plane anisotropy from nonlocal measurements in a four-terminal geometry, in which a pair of current leads is positioned far from the pair of voltage probes. Our rationale for resorting to such a method is the exponential sensitivity of the nonlocal signal to the anisotropy value. The exponential character of the nonlocal response follows from the seminal van der Pauw theorem,\cite{VanderPauw1958} which stipulates a relation between four-terminal resistance values $R_{ij,kl}=V_{ij}/I_{kl}$
obtained for the contacts $k$, $l$ and $i$,$j$ used as current leads and voltage probes, respectively. This  theorem connects the nonlocal resistance values for different combinations of voltage and current contacts in arbitrary-shape homogeneous 2D samples. For an isotropic conductor of the sheet conductivity $\sigma$ it reads  $e^{-\pi\sigma R_{12,34}}+e^{-\pi\sigma R_{23,41}}=1$, where the values $R_{ij,kl}$ depend on contact placement. Applied to long samples of dimensions $L\times W$, $L/W\gg 1$ with two pairs of contacts placed at opposite ends, the van der Pauw relations predict an exponential dependence of the nonlocal response on the aspect ratio 
\be
R_{\rm nl}\propto \rho e^{-\pi L/W},\quad \rho=\sigma^{-1}
.
\ee 
Below we extend this relation to anisotropic media using the {\it anamorphosis} (reshaping) theorem (see \emph{Appendix}).
\par
The anamorphosis theorem stipulates that the multi-terminal resistance values $R_{ij,kl}$ {\it remain unchanged} upon reshaping the sample via an anisotropic rescaling of the coordinates along the principal conductivity axes\cite{Price1972,Simon1999,Bierwagen2004,Mele2001,Wait1990}
\be\label{eq:reshaping}
x=\gamma^{1/2} x'
,\quad
y=\gamma^{-1/2} y'
,\quad 
\gamma=(\sigma_{xx}/\sigma_{yy})^{1/2}
,
\ee 
and simultaneously letting the conductivity of the reshaped system to be isotropic and equal  
\be\label{eq:sigma_ave}
\sigma'=(\sigma_{xx}\sigma_{yy})^{1/2}
.
\ee 
For long samples with the major axis parallel to the $x$ or $y$ axis the reshaping gives $L/W=\gamma^{\pm 1}L'/W'\gg 1$. The van der Pauw relation then predicts an exponential nonlocal response  altered by the anisotropy as
\be\label{eq:R_nl_LW}
R_{\text{nl}}^{(x)}\propto \rho' \exp\lp -\frac{\pi L}{W\gamma}\rp
,\quad
R_{\text{nl}}^{(y)}\propto \rho' \exp\lp-\frac{\pi L\gamma}{W}\rp
\ee
where we defined $\rho'=1/\sigma'$. Here $R_{\text{nl}}^{(x)}$ and $R_{\text{nl}}^{(y)}$ denote the nonlocal response for long samples aligned with the $x$ or $y$ axis, respectively. Our argument based on reshaping indicates robustness of the exponential $\gamma$ dependence. The order-one prefactors in Eq.\eqref{eq:R_nl_LW}, which in general are not universal, will be analyzed below for a strip geometry.
\par
\begin{figure}[t] %[!htb]
\includegraphics[width=\columnwidth]{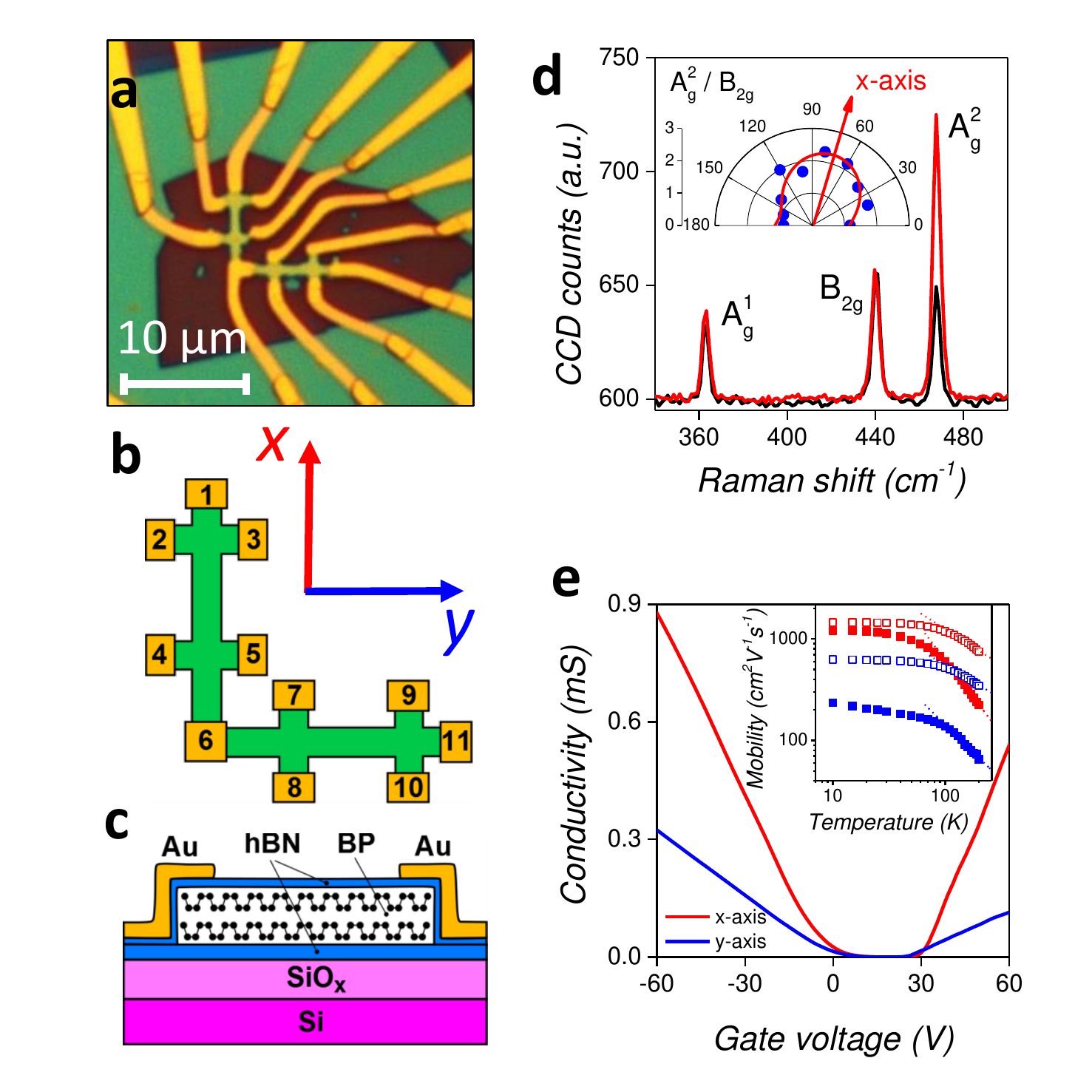}
\caption{Fabrication and characterization of crystal-axes-aligned devices of hBN-encapsulated few-layer phosphorene. a,b) Device micrograph and schematics: two orthogonal Hall bars oriented along the $x$ (easy) and $y$ (hard) crystal axes. c) Device cross-section: BP, encapsulated with hBN on SiO${}_{x}$/Si substrate, connected to gold contacts through tunnel barriers provided by the monolayer hBN. d) Polarization-dependent Raman spectra measured along two principal axes of hBN-encapsulated BP ($x$ axis, red; $y$ axis, black). Inset: Angle-resolved Raman signature $A_g^2 / B_{2g}$ used to determine the crystallographic orientation of a BP flake. 
e) Four-terminal conductivity of hBN-encapsulated BP \emph{vs.} gate voltage measured at $T = 5$K and bias voltage of $0.1$V ($x$ axis, red; $y$ axis, blue).  Inset: temperature dependence of the field-effect mobility for electrons (solid squares) and holes (empty squares). 
} 
\label{fig:fig1}
\end{figure}

The strong dependence on $\gamma$ suggests that nonlocal measurements afford a precision tool for determining the anisotropy. To exploit this opportunity we fabricated multiterminal Hall bars aligned with the $x$  and $y$ crystallographic axes of black phosphorous, see Fig.~\ref{fig:fig1}a--c. As discussed below, conductance anisotropy can then be extracted from the ratio of the nonlocal and local resistance. We found this approach to be reliable and robust despite a large variability in the nonlocal response values that could differ by orders of magnitude for  different crystallographic directions. 
\par
To better understand the impact of anisotropy on the nonlocal response, Eq.\eqref{eq:R_nl_LW}, we first consider an \emph{isotropic} conductor shaped into a Hall bar as in Fig.~\ref{fig:fig1}b. When current is passed between, say, contacts 2 and 3, it mostly flows straight across the device, with an exponentially small fraction flowing sideways, as the continuity equation requires that current does not drop to zero abruptly in the device channel. This produces stray currents which give rise to a voltage drop across remote contacts 4 and 5 (Fig.~\ref{fig:fig1}b). Analysis predicts a nonlocal resistance which decays exponentially with the separation $L$ between current leads and voltage probes:\cite{Abanin2011}
\be\label{eq:1}
R_{\text{nl}}\equiv \frac{V_{4,5}}{I_{2,3}} = \frac{2\rho}{\pi} \ln\coth\lp \frac{\pi L}{2W}\rp 
\approx \frac{4}{\pi} \rho \exp\lp -\frac{\pi L}{W}\rp 
\ee
where the asymptotic exponential form gives accuracy better than $0.1\%$ already for $L=W$. 
\par
A more complex behavior arises for an anisotropic conductor, where the anisotropic conductivity defines an easy  (high-conductivity) and a hard  (low-conductivity) direction for current flow. In this case, current applied along an easy direction is ``pinched'' into a relatively narrow path, which suppresses stray currents and  reduces nonlocal response. In contrast, the current applied along a hard direction tends to spread out due to an ``easy'' transverse flow short-circuiting the direct ``hard'' flow. In this case, enhanced stray currents lead to a large nonlocal response. The resulting suppression and enhancement of $R_{\text{nl}}$ in the two regimes is {\it exponential},  Eq.\eqref{eq:R_nl_LW}, with the corresponding exponents differing by the anisotropy factor $A=\gamma^2$ (see \emph{Appendix}).
\par
Next, we describe fabrication of the BP devices used in this work. Since few-layer black phosphorus is known to lack stability under ambient conditions,\cite{Island2015} we exfoliated BP crystals (purchased from HQ Graphene) in the inert atmosphere of a glove box. In order to protect BP from oxygen and moisture, crystals were encapsulated with hexagonal boron nitride (hBN), using the ``dry peel'' technique detailed elsewhere.\cite{Cao2015,Wang2013} Starting with freshly exfoliated BP on a polymer film, a thin ($\approx 10$ nm thickness) crystal  was identified and picked up with a monolayer hBN  on a polymer membrane.\cite{Kretinin2014} The resulting stack was then transferred onto a 30-80 nm hBN which served as an atomically flat substrate. The dry transfer method ensures an extremely clean interface between BP and hBN crystals and protects the former from degradation when air-sensitive few-layer BP is exposed to ambient conditions for many months\cite{Cao2015}. To avoid the degradation of BP surface during the fabrication procedures, the micromechanical exfoliation and transfer processes were performed in an argon environment provided by a glove box with levels of H${}_2$O and O${}_2$ below 0.1 ppm. The final assembly on 290 nm SiO${}_{\mathrm{x}}$/p-Si wafer was taken out of the glove box for subsequent microfabrication processing. 
\par
We used polarization-dependent Raman spectroscopy to establish the BP crystal axes orientation. Measurements were performed at room temperature in air using a confocal Witec spectrometer. Laser spot size of 0.5 $\mu$m was obtained by using a Nikon 100x objective. Samples were illuminated with green (2.41 eV) laser at 100 $\mu$W intensity. A high-density grating (1800 g/mm) was used to achieve energy resolution better than 2 cm${}^{-1}$. Angular control of the sample, relative to the laser polarization, was made with a rotating stage giving an uncertainty of about $\pm2$ degrees. Black phosphorus features three main Raman modes:\cite{Zhang2014} $A_g^1$ for the $z$ axis, out-of-plane phonons, $A_g^2$ for in-plane vibrations along the $x$ axis (armchair), and $B_{2g}$ for in-plane vibrations along the $y$ axis (zigzag). The amplitude of the two in-plane modes is polarization-dependent with $A_g^2$ intensity maximal for laser polarization along the  $x$ axis (armchair), see Fig.~\ref{fig:fig1}d. From the angular dependence of the normalized mode intensity $A_g^2 / B_{2g}$ we extracted the crystallographic orientation of BP crystals, as shown in Fig.~\ref{fig:fig1}d inset. 
\par
Subsequently, multiterminal Hall bar devices were defined via standard microfabrication techniques -- e-beam lithography and metal deposition followed by reactive ion etching -- with the device channel aligned with the $x$ or $y$ axis, Fig.~\ref{fig:fig1}a--c. Because the encapsulating hBN monolayer has relatively low ($< 1 $ k$\Omega\cdot\mu$m${}^2$) tunnel resistivity\cite{Britnell2012}, the electrodes (5nm Cr adhesion layer followed by 100 nm Au) were deposited directly onto it, Fig.~\ref{fig:fig1}c. Two devices (A and B) were studied in this work, each device comprising two orthogonal Hall bars defined along the $x$ and $y$ crystallographic axes. Devices A and B had channel length-to-width ratio $L/W$ of 3 and 2.5, respectively, with $W  = 1 \mu$m for both devices. Here we define the length $L$ as the distance between voltage probes (e.g. 2 and 4 or 8 and 10 in Fig.\ref{fig:fig1}b), the same length scale will be relevant for nonlocal measurements; see below. To characterize transport properties we measured four-terminal conductivity along both the $x$ and $y$ axes. DC conductivity was obtained by passing current through contact 1 to contact 11 and measuring voltage drop across, for example, contacts 2 and 4 for $\sigma_{xx}$ and contacts 8 and 10 for $\sigma_{yy}$, see Fig.~\ref{fig:fig1}b. Conductivity values were calculated as $\sigma_{xx} = R^{-1}_{xx}L/W$, $\sigma_{yy} = R^{-1}_{yy}L/W$.
\par  
Our field-effect transistor devices showed ambipolar behavior with somewhat enhanced hole-type transport (a likely result of p-type doping during crystal growth), as illustrated in Fig.~\ref{fig:fig1}e for device B. The overall device performance was comparable with that reported previously.\cite{Li2014} We extracted field-effect mobilities $\mu_{xx}$, $\mu_{yy}$ from the slopes $\delta\sigma / \delta V_g$  in Fig.~\ref{fig:fig1}e as $\mu=C^{-1}\delta\sigma / \delta V_g $. Here $C$ is the device channel-to-gate capacitance $C=\epsilon\epsilon_0/d$, where $d$ is the dielectric thickness. The capacitance was determined independently using Hall measurements. The field-effect mobilities plotted {\it vs.} temperature (see Fig.~\ref{fig:fig1}e inset) are nearly constant below 80K, decreasing as $T^{-\alpha}$ above 100K with $\alpha$ varying from 0.6 to 1.5 depending on the carrier type and crystallographic direction. We interpret the $T^{-\alpha}$ as the onset of electron-phonon scattering.
\par 
\begin{figure}[t] %[!htb]
\includegraphics[width=\columnwidth]{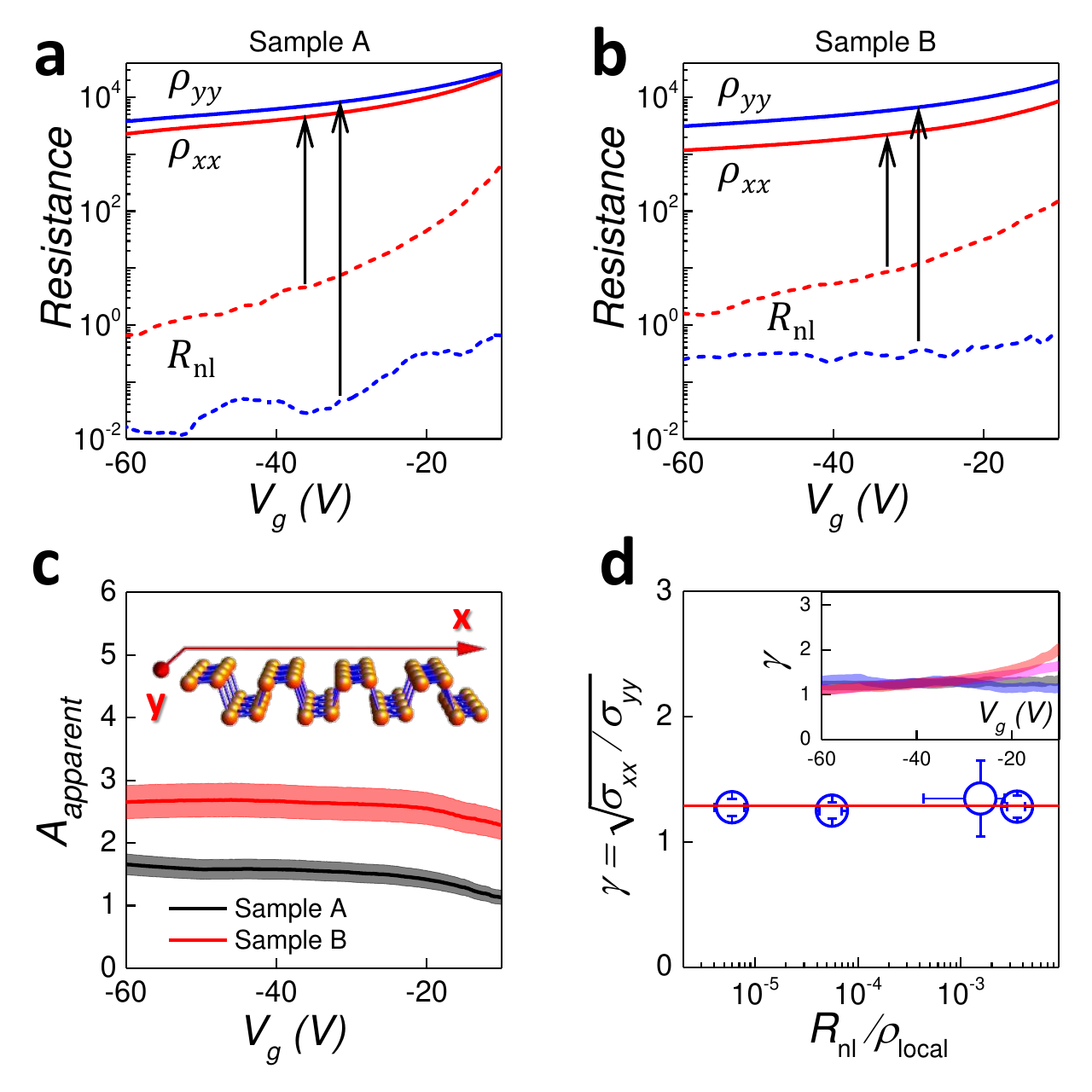}
\caption{Transport anisotropy of black phosphorus devices, $T = 5$K. a,b) Resistance \emph{vs.} gate voltage measured for samples A and B along principal crystallographic axes (red: $x$ axis, blue: $y$ axis, see panel c inset). Solid lines: local resistivity $\rho_{xx}$ and $\rho_{yy}$. Dashed lines: nonlocal resistance $R_{\text{nl}}$. Noise level $\approx 0.05 \Omega$. c) The apparent anisotropy $A_{\text{apparent}}=\sigma_{xx}/\sigma_{yy}$ measured as the ratio of the local (longitudinal) conductivities of the Hall bars aligned with the $x$ and $y$ axes. d) The anisotropy coefficient (averaged for $V_g$ varying from $-20$V to $-60$V) determined from the ratio of the nonlocal to local resistance, using $\rho_{\text{local}}=\rho_{xx}$ for $R_{\text{nl}}^{(x)}$ and $\rho_{\text{local}}=\rho_{yy}$ for $R_{\text{nl}}^{(y)}$. Inset shows the anisotropy coefficient \emph{vs.} gate voltage for both samples and both crystallographic axes (sample A: $x$ axis, red, $y$ axis, blue; sample B: $x$ axis, magenta, $y$ axis, black).   \label{fig:fig2}}
\end{figure}
At low temperatures (5-50K) the hole mobility values were approximately $ 1500$ and $500{\rm cm{}^2V{}^{-1}s{}^{-1}}$ for transport along the $x$ and $y$ axes, respectively. The corresponding electron mobility values were somewhat smaller, approximately $ 1200$ and $ 200$ cm${}^2$V${}^{-1}$s${}^{-1}$, respectively. The anisotropy of mobility was therefore between $2$ and $4$. Accounting for the Drude relation between mobility and conductivity, $\sigma = n e \mu$, yields the ratio $A=\sigma_{xx} / \sigma_{yy}$  as $1.5\pm0.1$ and $2.6\pm0.1$ for devices A and B, respectively (see Fig.~\ref{fig:fig2}c). Such a large spread in the apparent anisotropy values illustrates the pitfalls of this direct approach for accurate determination of anisotropy, Eq.\eqref{eq:A_defined}. The large spread arises even when the different devices used are made from the same crystal, and with a well-defined Hall bar geometry. 
\par
One can identify two main reasons for the direct method to lack precision\cite{Simon1999}. First, the value $A$ is obtained from conductivity measurements on physically different samples with somewhat different conductivity tensors. Second, even for well-defined geometry the aspect ratio might vary slightly from sample to sample. On the contrary, with our new method, both local and nonlocal measurements are done on the same device, thus we eliminate these two sources of inaccuracy. Further, the exponential sensitivity of the nonlocal response to the anisotropy value gives the nonlocal method a distinct advantage over local measurements in finding the true conductance anisotropy. 
\par
Next, we proceed to discuss the nonlocal resistance measurements. Due to the high contact resistance which is at least an order of magnitude larger than the channel resistance, we are able to reliably assess nonlocal resistances only in a region of $V_g$ from -10V to -60V, where the contact resistance is relatively small. We obtain nonlocal resistance by passing the current transverse the device channel (e.g., from contact 2 to 3 in Fig.~\ref{fig:fig1}b) and by measuring a voltage drop across the nearest pair of potential contacts (e.g., between contacts 4 and 5 in Fig.~\ref{fig:fig1}b). Figs.~\ref{fig:fig2}a and \ref{fig:fig2}b summarize results of our measurements for both local and nonlocal resistances using the two studied devices and both principal crystallographic axes. Nonlocal resistance was several orders of magnitude lower than the local resistance, and was also considerably lower along the $y$ axis than along the $x$ axis. Comparing panels a and b in Fig.~\ref{fig:fig2}, the difference in the nonlocal resistance values measured along the $x$ and $y$ axes is larger for the device with $L/W = 3$ than for the one with $L/W = 2.5$.  
\par
This behavior can be exemplified by four-terminal measurements in device B (in our discussion below we refer to the schematics shown in Fig.~\ref{fig:fig1}a,b). Nonlocal resistance measured along the $x$ axis, $R_{\text{nl}} = V_{4,5} / I_{2,3}$,  is a factor of 5-20 larger than the one measured along the $y$ axis, $R_{\text{nl}} = V_{9,10} / I_{7,8}$, see dashed red and blue lines in Fig.~\ref{fig:fig2}b. Indeed, as discussed above, since current between contacts 2 and 3 flows along the hard (low-conductivity) direction it tends to spread along the Hall bar channel (x-axis), resulting in a larger voltage drop across contacts 4 and 5. In contrast, current between contacts 7 and 8 flows along the easy (high-conductivity) direction, and is therefore pinched into a relatively narrow channel and spreads much less in the hard direction along the Hall bar channel (y-axis). This results in a much smaller voltage drop at remote contacts 9 and 10. 
\par
As discussed above, an anisotropic conductor can be treated as an isotropic one after an appropriate coordinate transformation (see \emph{Appendix}). This is accomplished by stretching/shrinking the device dimensions by a factor of $\gamma = \sqrt{\sigma_{xx}/\sigma_{yy}}$ and replacing the anisotropic conductivity tensor by an ``average conductivity value'',  Eq.\eqref{eq:sigma_ave}, giving
\be
R_{\text{nl}}^{(x)} =  \frac{4\gamma\rho_{xx}}{\pi}  \exp\lp -\frac{\pi L}{W\gamma} \rp
,\quad
R_{\text{nl}}^{(y)} =  \frac{4\rho_{yy}}{\pi\gamma}  \exp\lp -\frac{\pi L\gamma}{W} \rp
,	
\label{eq:2}
\ee
respectively. Hence, for a fixed geometry, the ratio of the nonlocal and local resistance depends only on the anisotropy coefficient $\gamma$. This enables us to extract $\gamma$ directly by combining the local and nonlocal measurements on a single Hall bar. Notably, Fig.~\ref{fig:fig2}d inset shows that $\gamma$ values are very similar for both devices and both crystallographic orientations of the Hall bars, and almost independent of gate voltage. Small deviations in $\gamma$ for  $V_g > -20$V stem from complications of current pathways as the Fermi level approaches band gap. Fig.~\ref{fig:fig2}d summarizes data from Figs.~\ref{fig:fig2}a and b and demonstrates the close agreement of anisotropy coefficient for both studied devices yielding $\gamma\equiv\sqrt{\sigma_{xx}/\sigma_{yy}} \approx 1.29 \pm 0.05$ and, hence, the anisotropy value $A\equiv \sigma_{xx}/\sigma_{yy} \approx 1.66 \pm 0.1$. 
\par
It is interesting to compare these results with the anisotropy in bulk BP. Anisotropy of Hall mobility measured at about 50K for p-type crystals was about $3.3$ (Ref.~\onlinecite{Akahama1983}), which is notably larger than the values found in our experiment as well as the values reported in Refs.~\onlinecite{Xia2014,Liu2014}. The twice smaller anisotropy of in-plane conductance of a few-layer phosphorene as compared to the bulk BP can be attributed to the surface scattering, which reduces the anisotropy of electron-phonon interaction.\cite{Luo2015} The anisotropic electron-phonon interactions as well as the anisotropic properties of acoustic phonons in black phosphorus could also lead to the anisotropy in Drude relaxation times $\tau_{yy} / \tau_{xx}$.\cite{Morita1989} The latter could explain the difference between the anisotropies of conductance and effective mass in BP. From previous measurements\cite{Narita1983} and theory\cite{Morita1986,Qiao2014} the ratio of effective masses along the $x$ and $y$ directions $m_{\rm eff}^{yy} / m_{\rm eff}^{xx}$ was estimated at $7.7\pm1.2$, which is much larger than the conductance anisotropy measured in this work. At constant carrier concentrations, the difference of the two anisotropies could be accounted for by the anisotropy $\tau_{yy} / \tau_{xx} \approx 4.4 \pm 1.1$. It would be interesting to find out how these anisotropies evolve with decreasing the number of layers down to monolayer phosphorene but, unfortunately, such encapsulated BP devices will require remote-controlled microfabrication in an inert atmosphere\cite{Cao2015} which, at present, appears to be very challenging.
\par
Lastly, we note that the nonlocal resistance observed in BP has purely classical explanation, thus differing it from nonlocal responses observed in systems with non-trivial band properties due to either spin-orbit coupling (spin Hall effect) or non-zero Berry curvature (anomalous/valley Hall effect). In the latter two cases, the nonlocal resistance is a manifestation of a combination of direct and inverse spin/anomalous Hall effects, induced by the presence of spin- or valley-polarized currents. These effects should play only a minor role (if any) in the nonlocal transport in BP because phosphorus is a light element and the spin-orbit terms do not contribute appreciably to BP band structure.\cite{Qiao2014} Moreover, topology of a band structure near the $\Gamma$ point of BP Brillouin zone is trivial (single valley),\cite{Rudenko2014} leading to zero Berry curvature and, thus, to the absence of topological currents.\cite{Lensky2014} 
\par
%%%%%%%%%%%%%%%               SUMMARY               %%%%%%%%%%%%%%%%%%%%%%%%%%%%%%%%%%%%%%%%%%%
Summing up, we have described a new approach to characterize transport anisotropy of layered nanomaterials which has distinct advantages over previously developed approaches. Our method leverages on novel measurement schemes (nonlocal response) through which the effect of anisotropy can be sharply enhanced and accessed directly without resorting to traditional angle-resolved approaches. Namely, our method utilizes the combination of local and nonlocal transport measurements and relies on the exponential sensitivity of nonlocal resistance to the anisotropy coefficient. Using this method we find that the anisotropy of in-plane conductance $\sigma_{xx}/\sigma_{yy}$ for thin BP crystals amounts to $\approx 1.66 \pm 0.1$. This method can easily be adapted to other potentially anisotropic 2D crystals (e.g. ReSe${}_2$ and ReS${}_2$) as well as for 2D materials and heterostructures where in-plane anisotropy might be induced via strain or charge density waves.
\begin{acknowledgments}
This work was supported by European Research Council Synergy Grant Hetero2D, EC-FET European Graphene Flagship, The Royal Society, U.S. Army, Engineering and Physical Sciences Research Council (UK), the Leverhulme Trust (UK), U.S. Office of Naval Research, U.S. Defence Threat Reduction Agency, U.S. Air Force Office of Scientific Research. L.S.L. acknowledges support of the Center for Integrated Quantum Materials (CIQM) under NSF award 1231319. 
\end{acknowledgments}

\appendix
\section{The Anamorphosis Theorem\label{sec:appendix1}}
Here we derive a simple and general relation that connects multi-terminal resistance values for an anisotropic conductor and those for a reshaped isotropic conductor of conductivity $\sigma'=(\sigma _{xx}\sigma _{yy})^{1/2}$. This relation, thereafter referred to as {\it the anamorphosis (reshaping) theorem},  is valid for homogeneous samples of an arbitrary shape with the contacts placed in an arbitrary fashion, either at the sample boundary or in the interior. The theorem stipulates that the multiterminal resistance values $R_{ij,kl}$, defined as the potential difference on the contacts $i$ and $j$ induced by a unit current fed through contacts $k$ and $l$, remains unchanged upon sample reshaping:
\be\label{eq:R'=R}
R'_{ij,kl}=R_{ij,kl}
%,\quad
%\sigma'=(\sigma _{xx}\sigma _{yy})^{1/2}
.
\ee
Here $R_{ij,kl}=V_{ij}/I_{kl}$, $R'_{ij,kl}=V'_{ij}/I'_{kl}$ where the unprimed and primed quantities correspond to the original and reshaped sample, respectively. Eq.\eqref{eq:R'=R} provides a simple and model-independent relation between different transport measurement results. Unlike the van der Pauw theorem\cite{VanderPauw1958} which connects multiterminal resistance values obtained for the same sample by a permutation of current and voltage contacts, the result in Eq.\eqref{eq:R'=R} relates the values obtained for different samples albeit with no contact permutation.

To establish the result in Eq.\eqref{eq:R'=R}, we recall that  the conductivity of an anisotropic 2D conductor, in the absence of magnetic field, is described by a second rank tensor. Without loss of generality, we can pick the axes of our coordinate system to be aligned with the principal conductivity axes, writing
\be
\sigma = \left[\begin{array}{cc}\sigma _{xx} & 0 \\ 0 & \sigma _{yy} \end{array}\right]
.
\ee
%where, without loss of generality, we picked the axes of our coordinate system to be aligned with the principal conductivity axes. Here we will be concerned with the ohmic transport governed by the equations
In these coordinates, the equations describing ohmic transport read
\be\label{eq:j-E}
j _{x}=\sigma_{xx} E _{x}
,\quad
j _{y}=\sigma_{yy} E _{y}
\ee
where $\vec E=(E _{x},E _{y})=-\nabla\phi$. 
Here the current $\vec j$, electric field $\vec E$, and the potential $\phi$ spatial dependence are obtained from a solution of a boundary value problem posed through the current continuity relation 
\be\label{eq:div j=0}
\p_i j_i=0
% \frac{\p j _{x}}{\p x}+\frac{\p j _{y}}{\p y}=0
.
\ee
This equation, which holds inside the sample material, must be solved together with the boundary condition specifying fixed potential values at the contacts $\phi_i$, $i=1...4$ and the tangential-current condition $\vec l\times \vec j=0$ at the nonconducting parts of the boundary (here $\vec l$ is a vector tangential to the boundary).

In order to map the anisotropic problem to an isotropic one we introduce an anisotropic scaling transformation
\be\label{eq:rescaling}
x =\eta x',\quad
y=\eta^{-1} y'
,\quad
\eta=(\sigma _{xx}/\sigma _{yy})^{1/4} 
\ee
We note that under this transformation 
the continuity relation given in 
Eq.\eqref{eq:div j=0} continues to hold provided the current components are redefined as
\be
j _{x}=\eta j' _{x}
,\quad
j _{y}=\eta^{-1} j' _{y}
.
\ee
Crucially, $\vec j'$ satisfies the current-field relations for an isotropic conductor of conductivity value $\sigma'$:  % given in Eq. 3 in the main text:
% $\sigma'=(\sigma _{xx}\sigma _{yy})^{1/2}$:
%LL Next, the current-field relations \eqref{eq:j-E} transform as
\be
j' _{x}= \sigma' E' _{x}
,\quad
j' _{y}= \sigma' E' _{y}
,\quad
\vec E'=(E' _{x},E' _{y})=-\nabla'\phi
\ee
where 
%we defined $\sigma'=(\sigma _{xx}\sigma _{yy})^{1/2}$ (here 
$\nabla'=(\eta^{-1}\p_{x'},\eta \p_{y'})$.

Next, it is straightforward to check that the reshaping transformation in Eq.\eqref{eq:rescaling} leaves the tangential-current boundary condition $\vec l\times \vec j=0$ unchanged. Indeed, upon rescaling in Eq.\eqref{eq:rescaling} the tangential vector changes as $\vec l'=(\eta l _{x}, \eta^{-1}l _{y})$, yielding $\vec l'\times \vec j'=0$ (our tangential vectors $\vec l$ and $\vec l'$ are not normalized.) We therefore conclude that new transport equations are identical to those for an isotropic material of conductivity $\sigma'$. 

Lastly, we note that the definitions of voltages and currents at the contacts remain unchanged upon rescaling: $V'_k=V_k$ and 
\be
I'_k=\int d \vec l'\times\vec j'=\int d\vec l\times\vec j  =I_k
,
\ee 
where the integral is taken along the boundary of the $k$th contact before and after rescaling. As a result, the multi-terminal resistance  values for the transformed geometry, in which both the sample and the contacts are reshaped, are identical to those in the original anisotropic sample:
% lso remain unchanged:
%can be rewritten as
\[
R'_{ij,kl}=\frac{V'_{ij}}{I'_{kl}}=\frac{V_{ij}}{I_{kl}}=R_{ij,kl}
,
\]
which proves the reshaping theorem, Eq.\eqref{eq:R'=R}.

As a sanity check, consider a conducting strip of width $W$ parallel to the $x$ axis which carries constant uniform current $I$. Voltage drop between probes on the strip edge separated by a distance $L$ is then given by $V=(L/W\sigma_{xx})I$. After the reshaping  transformation, Eq.\eqref{eq:rescaling}, the new strip width and the new distance between voltage probes are $W'=\eta W$ and $L'=\eta^{-1} L$. Accounting for the isotropic conductivity $\sigma'$ of the reshaped conductor we find that voltage remains unchanged upon reshaping:
\be
V'=\frac{L'}{W'\sigma'}I=\frac{L}{W\sigma_{xx}}I=V
,
\ee
which is in full accord with our anamorphosis theorem.
%, Eq.\eqref{eq:R'=R}.

\begin{figure}[!htb]
 \includegraphics[width=0.8\columnwidth]{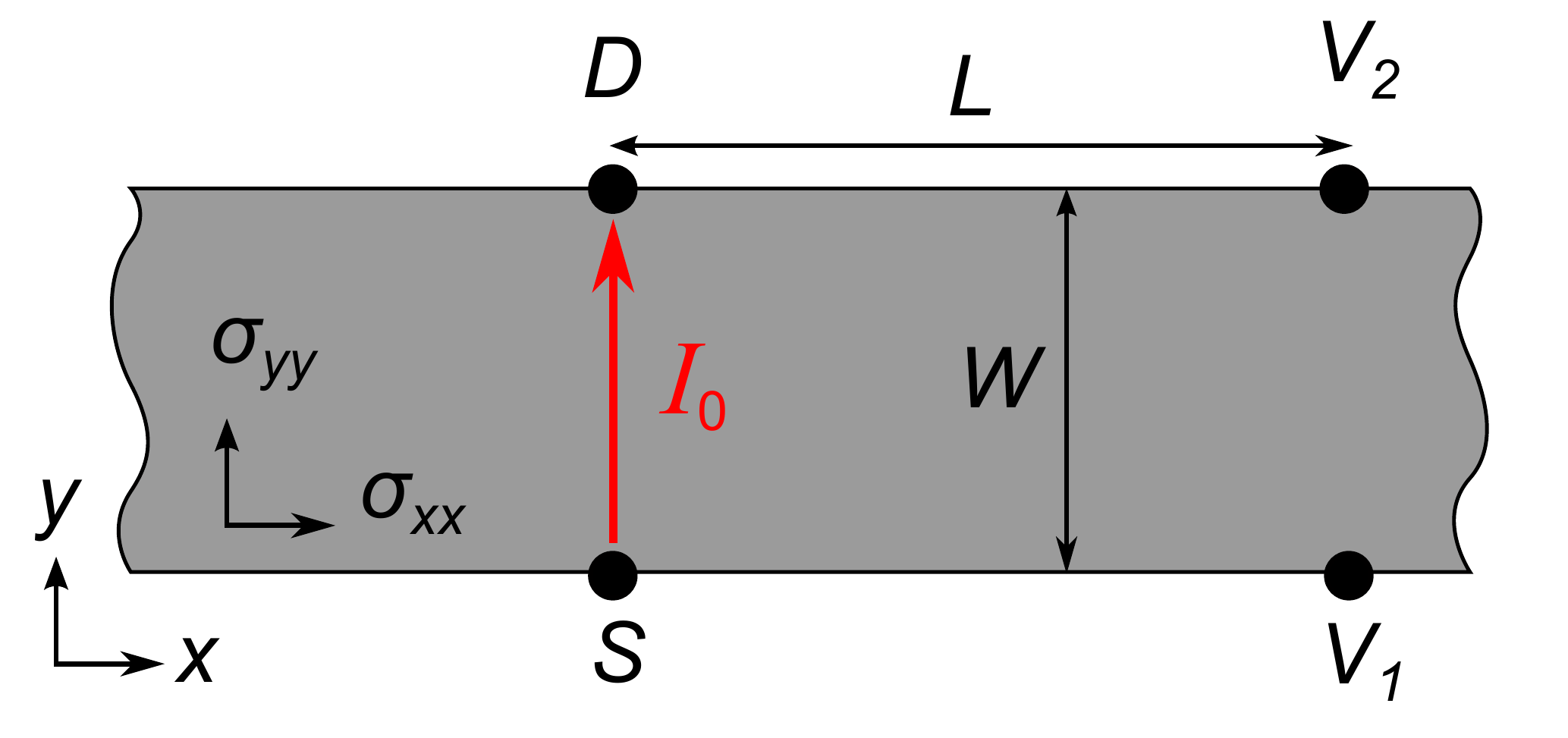}
 \caption{Schematics of anisotropic conductor modelled as infinite strip of width $W$. Principal axes of conductivity tensor $\lp\sigma_{xx},\sigma_{yy}\rp$ are aligned with the $x$ and $y$ axes. Current $I $ is applied between source ($S$) and drain ($D$) electrodes. Voltage probes $V_1$ and $V_2$ are placed at a distance $L$ away from $S-D$ pair.}
	\label{SI_fig1}
 \end{figure}

Next, we consider currents and potentials for a nonlocal measurement in the same strip. The corresponding contact placement is illustrated in Fig.\ref{SI_fig1}. Performing the reshaping transformation to a strip with an isotropic conductivity $\sigma'$, as above, we arrive at a problem which has been solved elsewhere.\cite{Abanin2011} In this case the nonlocal voltage induced at a distance $L'$ from current leads is given by
\be\label{eq:R_NL_2011}
V=\frac{\rho'}{\pi}\ln\lp \frac{\cosh\pi\frac{L'}{W'}+1}{\cosh\pi\frac{L'}{W'}-1} \rp I
\ee
where $\rho'=1/\sigma'$. This relation can also be written as
\be\label{eq:R_NL_coth}
V=\frac{2\rho' I}{\pi}\ln\coth\lp\frac{\pi L'}{2W'}\rp
.
\ee
For $L'\gg W'$ this gives
\be
V\approx \frac{4\rho'}{\pi}e^{-\pi\frac{L'}{W'}} I
.
\ee
These expressions, according to the relation  \eqref{eq:R'=R} proven above, 
%our {\it anamorphosis} theorem Eq.\eqref{eq:R'=R} proven above, 
also describe the nonlocal response in an anisotropic system, provided that the corresponding  lengthscales are related via $W'=\eta W$,  $L'=\eta^{-1}L$. This provides a derivation of the expression for nonlocal resistance used in the main text. 

%%%%%%%%%%%%%%%%%%%%%%%%%%%%%%%%%%%%%%%%%%%%%%%%%%%%%%%%%%%%%%%%%%%%%%%%%%%%%%%%%%%%%%%%%%%%%%%%%%%%%%%%%%%%%

\section{Direct derivation of the nonlocal response for a strip geometry\label{sec:appendix2}}

Here we derive the nonlocal response for a strip geometry by a direct method that does not rely on the reshaping theorem. We proceed by combining the current-field relations, Eq.\eqref{eq:j-E}, with the continuity equation $\p_i j_i=0$, which gives an anisotropic Laplace's equation for the potential 
% $\varphi(x,y)$
%
%For an anisotropic 2D semiconductor, the conductivity in the absence of magnetic field is the second rank tensor $\bar{\bar{\sigma}} = \left[\begin{array}{cc}\sigma_{xx} & 0 \\ 0 & \sigma_{yy} \end{array}\right]$.
%
% 
%Following the procedure described elsewhere\cite{Abanin2011} we consider an infinite strip of anisotropic conductor as in Fig.~\ref{SI_fig1}. The current-field relations are
%\be
%\mathbf{j} = \bar{\bar{\sigma}} \cdot \mathbf{E} =  \left[ \begin{array}{cc} \sigma_{xx} & 0 \\ 0 & \sigma_{yy}  \end{array} \right] \cdot \left[ \begin{array}{cc} E_{xx} \\ E_{yy} \end{array} \right]= \left[ \begin{array}{cc} -\sigma_{xx} \partial \varphi / \partial x \\ -\sigma_{yy} \partial \varphi / \partial y \end{array} \right] 
%\label{app:eq:IV}
%\ee
%Here the current and electric field vectors $\mathbf{j}$ and $\mathbf{E}$, as well as the potential $\varphi$, are fields obtained from a solution of a boundary value problem posed through the current continuity relation inside the material: 
\be\label{eqn:laplace_anisotropic}
\lp \sigma_{xx}\frac{\partial^2 }{\partial x^2}+\sigma_{yy}\frac{\partial^2 }{\partial y^2} \rp \varphi(x,y)= 0
% \label{app:eq:cont}
\ee
This equation must be solved in a strip 
\be\label{eq:strip}
-\infty<x<\infty
,\quad
-y_0<y<y_0
,\quad
y_0=W/2
\ee
(see  Fig.\ref{SI_fig1}) with the
source and drain current leads positioned at $(x,y)=(0,\pm y_0)$ and potential probes located at a distance $L$ on the opposite edges of the strip, $(x,y)=\left(L,\pm y_0\right)$. 
 
%From the extension of van der Pauw theorem\cite{VanderPauw1958} for anisotropic medium  it follows that the equation of continuity in an anisotropic medium can be transformed to canonical (isotropic) form via a  scaling transformation of coordinates.\cite{Price1972,Simon1999,Bierwagen2004} We choose $x=X\sqrt[4]{\sigma_{xx}/\sigma_{yy}}$ and $y=Y\sqrt[4]{\sigma_{yy}/\sigma_{xx}}$. This transforms anisotropic conductor to isotropic one with new dimensions $L'\times W'$ and with isotropic conductivity defined as $\sigma'=\sqrt{\sigma_{xx}\sigma_{yy}}$. Boundary conditions  are now $-\sigma'\frac{\partial \varphi(X,Y)}{\partial Y}\bigg|_{y=\pm\frac{W'}{2}} =I \delta(x)$. Eq.~\ref{app:eq:cont} then satisfies the Laplace's equation: 
%\be
%\frac{\partial^2 \varphi(X,Y)}{\partial X^2}+\frac{\partial^2 \varphi(X,Y)}{\partial Y^2} = 0
%\label{eqn:laplace}
%\ee

Accounting for translational invariance of the strip \eqref{eq:strip}, we perform Fourier transform via 
$\tilde\phi(k,y)=\int dx e^{-ikx}\phi(x,y)$, 
converting Eq.\eqref{eqn:laplace_anisotropic} into a a second-order ordinary differential equation 
\be
\lp -\kappa^2+\frac{\partial^2 }{\partial y^2}\rp \tilde\phi (k,y) = 0
,\quad
\kappa=\gamma k
,
\label{eqn:ode}
\ee
where $\gamma=\lp \sigma_{xx}/\sigma_{yy}\rp^{1/2}$ as above. 
Eq.\eqref{eqn:ode} has a general solution
\be
\tilde\phi(k,y) = A_k e^{\kappa y}+B_k e^{-\kappa y}
%,\quad
%\kappa=\gamma k
. 
\label{app:eq:sol1}
\ee
Current entering and leaving the system through the source and drain leads is modeled by a Dirac delta function in the boundary conditions for the current density normal component,  $j_{y} (x) =I\delta(x)$ at $y=\pm y_0$. % for the continuity equation. 
Writing these relations as 
% the boundary conditions at $y=\pm y_0$ as
$\partial \tilde\phi_k /\partial y =  -I/\sigma_{yy}$
and solving for $A_k$, $B_k$, gives 
% $A(k)$ and $B(k)$ for Eq.~\ref{app:eq:sol1}
\be
A_k=
-B_k=-\frac{I }{2\kappa\sigma_{yy}\cosh \kappa y_0 }	
\ee
Plugging these values in Eq.\eqref{app:eq:sol1} we find
% Then the solution to Eq.~\ref{eqn:ode} is
\be
\tilde\phi(k,y) = -\frac{I \sinh{\kappa y}}{k\sigma'\cosh {\kappa y_0} }		
\ee
where we used the relation $\kappa\sigma_{yy}=k\sigma'$. 
Inverting Fourier transform yields a solution to the original problem 
\be
\varphi(x,y) = -I \int_{-\infty}^{\infty}\frac{dk}{2\pi\sigma'}\frac{e^{-ikx}\sinh{\kappa y}}{k\cosh {\kappa y_0} }
\ee
For the potential difference across the strip at a distance $x=L$ from current leads, $V(L)=\varphi(L,-y_0)-\varphi(L,y_0)$, this gives
\be
V(L) = \frac{I }{\pi\sigma'}\int_{-\infty}^{\infty}\frac{e^{-ikL}\tanh(\kappa y_0)}{k} dk	
.	
\ee
This expression can be evaluated with the help of a contour integral
%LL where the integration path runs along $-\infty<z<+\infty$, returning back at infinity, gives
\be
\oint e^{2i\alpha z}\frac{\tanh z}{z}dz=4\sum_{n\ge 0}\frac{e^{-\pi(2n+1)\alpha}}{2n+1}
=2\ln\frac{1+e^{-\pi\alpha}}{1-e^{-\pi\alpha}}
\ee
where $\alpha>0$ and the integration path runs along $-\infty<z<+\infty$, returning back at infinity. Applied to $V(L)$, it yields 
% the result
%relation identical to that derived above
\be
V(L) = \frac{2I }{\pi\sigma'}\ln\coth\lp\frac{\pi L}{2\gamma W}\rp
,
%\left[\frac{\cosh\pi\frac{L}{\gamma W}+1}{\cosh\pi\frac{L}{\gamma W}-1}\right]		
\ee
which is in agreement with the result noted above, Eq.\eqref{eq:R_NL_coth}. 
The above expression for nonlocal response is identical to that in Eq.\eqref{eq:R_NL_2011}. 
For $L\gg W$ this gives 
\be
R_{\text{nl}}  \approx \frac{4}{\pi\sigma'}\exp\lp -\frac{\pi L}{\gamma W}\rp
.
\ee
%Changing back to the initial coordinate system, $R_{\text{nl}} = \Delta \varphi(L,\mp\frac{W}{2}) / I $ is then 
%\be
%R_{\text{nl}} \approx \frac{4}{\pi\sigma'}\exp\left[-\pi\frac{L\sqrt[4]{\sigma_{yy}/\sigma_{xx}}}{W\sqrt[4]{\sigma_{xx}/\sigma_{yy}}}\right] = \frac{4}{\pi\sigma'}e^{-\pi\frac{L}{W}\gamma^{-1}}		
%\ee
%where $\gamma = \sqrt{\sigma_{xx}/\sigma_{yy}}$. 
From this result we obtain the ratio of nonlocal resistance $R_{\text{nl}}$, measured for a strip aligned with the $x$ ($y$) axis, and the corresponding resistivity $\rho_{xx}$ ($\rho_{yy}$): 
\be
\frac{R_{\text{nl}}^{(x)}}{\rho_{xx}} = \frac{4 \gamma}{\pi} \exp\lp-\frac{\pi L}{W \gamma } \rp	
,\quad
\frac{R_{\text{nl}}^{(y)}}{\rho_{yy}} = \frac{4}{\pi\gamma} \exp\lp-\frac{ \pi L \gamma}{W} \rp	
.
\ee
With the quantities on the left hand side known from measurements, each of these nonlinear equations can  be solved individually to obtain the $\gamma$ value. This can be conveniently written
 in terms of the so-called Lambert $\mathbb{W}_0$ function:
\be\label{eq:Lambert}
\gamma = \left[\frac{W}{\pi L}\mathbb{W}_0\left(\frac{4L}{W}\frac{\rho_{xx}}{R_{\text{nl}}}\right)\right]^{-1}
, \quad
\gamma =\frac{W}{\pi L}\mathbb{W}_0\left(\frac{4L}{W}\frac{\rho_{yy}}{R_{\text{nl}}}\right)
\ee
%\addLL{LL: I think you are using a different function, namely the inverse of $y=xe^{-x}$. That, if so desired, can be written as $x=-\mathbb{W}_0(-y)$.}
The  function $\mathbb{W}_0$ is defined as the inverse of the function $f(x) = x e^x$ in the domain $x>0$ (the subscript $0$ denotes the principal real branch of the function). The Lambert function is easily accessible from most of mathematics software products, including OriginPro\footnote{in OriginPro the Lambert 
%LL $\mathbb{W}_0$ 
function can be accessed as described in the folloowing link: http://www.originlab.com/doc/LabTalk/ref/LambertW-func}, MATLAB\footnote{Good reference on
% introduction and explanation of 
how to use Lambert 
%LL$\mathbb{W}$ 
function in MATLAB products is provided here: http://uk.mathworks.com/help/symbolic/lambertw.html} and Wolfram Mathematica\footnote{For a description of $\mathbb{W}_0$ and its use in Mathematica products look at http://mathworld.wolfram.com/LambertW-Function.html}.
This procedure gives two independent estimates of the anisotropy value $A=\sigma_{xx}/\sigma_{yy}$. Notably, despite the fact that in our measurements $R_{\text{nl}}^{(x)}$ and $R_{\text{nl}}^{(y)}$ typically differ by orders of magnitude, the relations \eqref{eq:Lambert} yield nearly identical $\gamma$ values for the Hall bars oriented along different crystal axes on the same device. Similar values are also found for measurements on different devices.  

\bibliographystyle{artemnl}
\bibliography{BPliterature}

\end{document}